\documentclass[preprint,12pt]{elsarticle}
\usepackage{graphicx}

\usepackage{amssymb}

\journal{Physica A}

\begin{document}

\begin{frontmatter}



\title{Splitting of the ground state manifold of classical Heisenberg spins as couplings are varied}


\author{Samarth Chandra}

\address {Department of Theoretical Physics, Tata Institute of Fundamental Research, Homi Bhabha Road, Mumbai 400 005, India \fnref{currentaddress}}

\fntext[currentaddress]{Currently at: Raman Research Institute, C. V. Raman Avenue, Sadashivanagar, Bangalore-560080, India}

\ead{samarth@rri.res.in}

\begin{abstract}

 We construct clusters of classical Heisenberg spins with two spin $\vec{S}_i . \vec{S}_j$-type interactions for which the ground state manifold consists of disconnected pieces. We extend the construction to lattices and couplings for which the ground state manifold splits into an exponentially large number of disconnected pieces at a sharp point as the interaction strengths are varied with respect to each other. In one such lattice we construct, the number of disconnected pieces in the ground state manifold can be counted exactly.

\end{abstract}


\end{frontmatter}



 Classical as well as quantum spin systems are of basic importance in condensed matter as well as molecular physics \cite{label1}. Spin systems with competing interactions give rise to fascinating phenomena like spin glasses \cite{label2},\cite{label3}. In this context, classical spin systems with hamiltonians of the form $\mathcal{H} = - \sum_{i<j} J_{ij} \vec{S}_i . \vec{S}_j$ have been widely studied, various lattices being obtained by setting specific $J_{ij}$'s to zero. In the case of single molecule magnets, there are a few spins in the molecule. If the value of a spin is high it behaves similar to a classical spin. In this context, clusters of few classical Heisenberg spins with $\vec{S}_i . \vec{S}_j$-type interactions are of interest. The ground states of the system are important because they are indicative of the properties of the system at low temperatures.

 An important question that arises is whether the ground state manifold of such systems can have disconnected pieces in it or will it always be a single connected piece ? By disconnected pieces in the ground state manifold we mean that to go continuously from a state in one piece of the manifold to a state in a different piece of the manifold the system has to \emph{necessarily} go to excited states in between. If the ground state manifold has disconnected pieces, its dynamics will be significantly different from what it would be otherwise. To give an example, a bistable system with two disconnected pieces in the ground state manifold may display the phenomenon of stochastic resonance \cite{label4}.

 For the case where the vectors $\vec{S}_i$ are discrete Ising spins the state space itself is discrete. For XY-spins on a triangular lattice, interacting antiferromagnetically, the ground state manifold has two disconnected pieces corresponding to whether the spins on any given elementary triangle are arranged at $120^o$ to eachother in a clockwise or in an anticlockwise manner.  For Heisenberg spins with the most commonly studied interactions of two spin $\vec{S}_i . \vec{S}_j$-type it is not known whether the ground state manifold can have disconnected pieces or not.

 In this article we construct clusters, and also translationally invariant lattices, with Heisenberg spins placed on them, see figure 1 and 2, such that for low values of the tuning parameter, $x$, there is only one ground state manifold. However as $x$ is increased, at a sharp point, the ground state manifold of the cluster splits into two disconnected pieces while that of the lattices into an exponential number of pieces. These pieces of the ground state manifold continuously move apart as the tuning parameter is increased further. We then construct another lattice with similar properties but for which we can enumerate the number of disconnected pieces in the ground state manifold exactly.

\begin{figure}
\centering
\includegraphics[width=.8\columnwidth,angle=0]{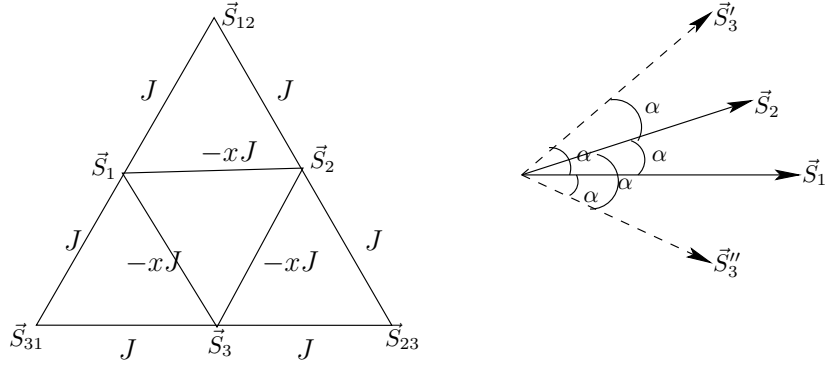}
\caption{The six-spin cluster and its ground state. The spins $\vec{S}_1$ and $\vec{S}_2$ form an angle of $\alpha$ with each other. $\vec{S}_3'$ rises above the plane of $\vec{S}_1$ and $\vec{S}_2$ forming an angle of $\alpha$ with both of them. $\vec{S}_3''$ goes below the plane of $\vec{S}_1$ and $\vec{S}_2$ forming an angle of $\alpha$ with both of them.}
\label{cluster}
\end{figure}

\begin{figure}
\centering
\includegraphics[width=1.1\columnwidth,angle=0]{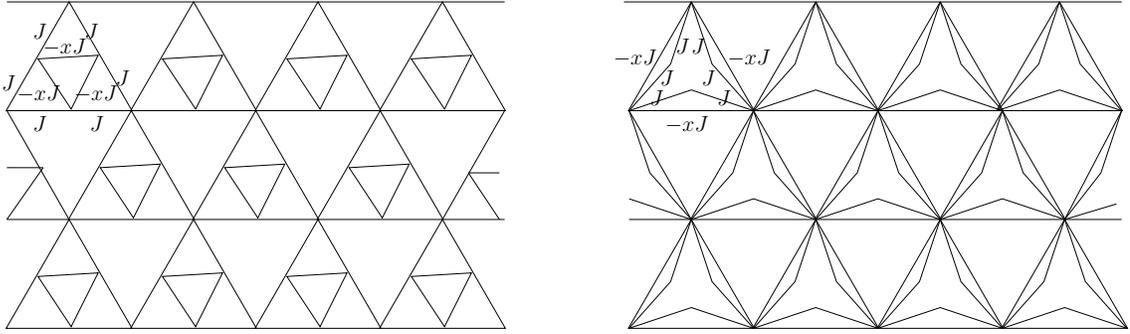}
\caption{Translationally invariant hamiltonians which show a splitting of the ground state manifold into an exponentially large number of pieces as $x$ is tuned.}
\label{lat}
\end{figure}

 The cluster of figure 1 and the lattices of figure 2 have been formed out of piecing together triangles of the form shown in figure 3, so we analyse the triangle of spins in figure 3 first.

\begin{figure}
\centering
\includegraphics[width=0.6\columnwidth,angle=0]{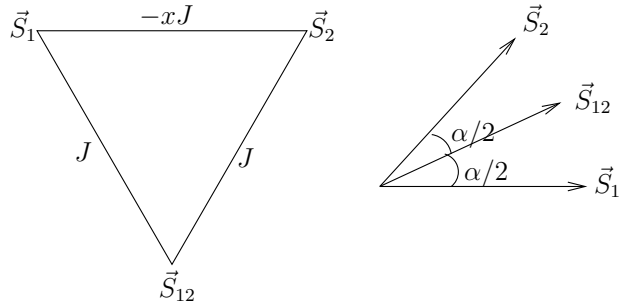}
\caption{An $S_2$ subgraph between spins $\vec{S}_1$ and $\vec{S}_2$ and its ground state}
\label{tri}
\end{figure}

 For three Heisenberg spins with two-spin $\vec{S}_i . \vec{S}_j$-type interactions, there always exists a ground state with all the three spins coplanar \cite{label5}. The hamiltonian of the spins of figure 3 is $\mathcal{H}_{tri} = x J \vec{S}_1 . \vec{S}_2 - J \vec{S}_{12} . (\vec{S}_1 + \vec{S}_2)$. Clearly, for any value of $x$, in the ground state, the spin $\vec{S}_{12}$ will point along  the direction $(\vec{S}_1 + \vec{S}_2)$. For $x < \frac{1}{2}$ the only minimum of energy is at $\alpha = 0$ while for $x > \frac{1}{2}$ the minimum occurs at $\alpha = 2 \arccos{\left(\frac{1}{2x}\right)}$. Thus for $x<\frac{1}{2}$ the spins are parallel to eachother in the ground state while for $x>\frac{1}{2}$ the spins $\vec{S}_1$ and $\vec{S}_2$ make an angle of $\left(2 \arccos{\left(\frac{1}{2x}\right)}\right)$ and $\vec{S}_{12}$ lies along their angle bisector. The angle $\alpha$ increases continuously to $\pi$ as x increases.

Now consider what happens when such triangles are put together.

 The six spin cluster shown in figure 1, is made of three triangles of the type shown in figure 3. For any value of $x$ the energy of the three triangles gets individually minimised when the three spins, $\vec{S}_1$, $\vec{S}_2$ and $\vec{S}_3$ each make an angle of $\alpha$ ($\alpha=0$ if $x<\frac{1}{2}$, and $\alpha = 2 \arccos{\left(\frac{1}{2x}\right)}$ if $x>\frac{1}{2}$) with respect to eachother and the rest of the three spins are at the respective angle bisectors. Thus given the positions of $\vec{S}_1$ and $\vec{S}_2$, making an angle of $\alpha$ between them, there are two different ground states of the system as shown in the figure. These two states being of different chirality can not be rigidly rotated into eachother. To continuously deform one of them into the other one has to necessarily change the angles between the spins, which raises the energy. Thus there are two disconnected pieces of the ground state manifold, obtained by the rigid rotation of each of these ground states.

 The lattice on the left in figure 2 can be obtained by putting together six-spin clusters of figure 1, joined at the spins $\vec{S}_{12}$, $\vec{S}_{23}$ and $\vec{S}_{31}$ while the lattice on the right in figure 2 can be obtained by joining at spins $\vec{S}_{1}$, $\vec{S}_{2}$ and $\vec{S}_{3}$. Arguing as above, it is easy to see that for $x<\frac{1}{2}$ there is a single connected ground state manifold. As we raise the value of $x$, at $x=\frac{1}{2}$, the ground state manifold splits into a number of disconnected pieces which is exponential in $N$, where $N$ is the total number of spins in the lattice. To see this, for instance, the lattice on the right in figure 2 is a decorated triangular lattice. The underlying triangular lattice is obviously tripartite with three sublattices, say A,B and C. Let the spins of the sublattice A all point in the same direction and those of sublattice B all in another direction which makes an angle $\alpha$ to those of A. Then the $(N/6)$ spins of sublattice C can independently point in either of the two directions making an angle $\alpha$ with A as well as B spins thus obtaining a ground state (the rest of the spins point along the respective angle bisectors). Thus the number of pieces in the ground state manifold is atleast $2^{(N/6)}$ and, it can be easily seen, is less than $2^{(N/2)}$, thus completing the proof. These pieces continuously move apart as $x$ is increased.

 We now describe the construction of a lattice of \emph{edge} sharing triangles for which we can count the number of disconnected pieces in the ground state manifold exactly.

 Take a three coordinated bethe lattice, or any of its connected subgraphs. With each vertex associate a triangle such that its three edges are respectively perpendicular to the three bonds coming out of the vertex, the triangles of two neighbouring vertices sharing the edge perpendicular to the bond joining them, with the corresponding vertices identified. An example is shown in figure 4. Now each edge of this new graph is itself taken to be an $S_2$-subgraph shown in figure 3. For $x<\frac{1}{2}$, there is only one connected ground state manifold. For $x>\frac{1}{2}$, there are exactly $2^T$ disconnected pieces of the ground state manifold, $T$ being the number of vertices in the subgraph of the Bethe lattice. To see this, pick any vertex in the original subgraph of the bethe lattice. The three spins on the elementary triangle around it can be arranged in two ways of different chirality at an angle of $\alpha$ to each other (the internal spins of the $S_2$ subgraphs get fixed uniquely). Now move along the original subgraph of the bethe lattice one edge at a time. The spin at each new vertex of the elementary triangle thus encountered can be arranged in two ways of different chirality maintaining an angle $\alpha$ with the two other spins of this elementary triangle, thus completing the proof.

\begin{figure}
\centering
\includegraphics[width=0.5\columnwidth,angle=0]{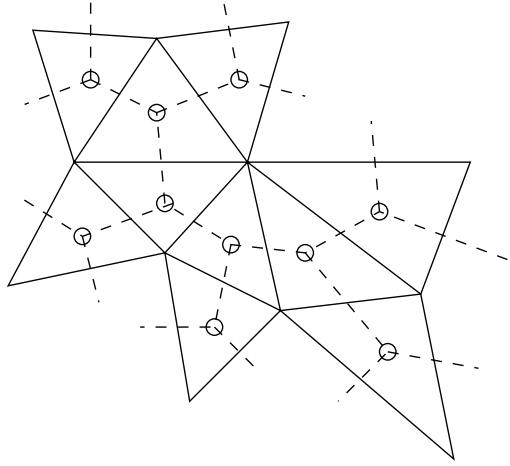}
\caption{A subgraph of the Bethe lattice and the construction of the corresponding edge-sharing triangle graph.}
\label{corner_tri}
\end{figure}

 Let us further study the distribution of the various disconnected pieces of the ground state manifold of this lattice. If we change our choice of orientation of any spin, say $\vec{S}_1$, in the bulk of the lattice, all the spins from there to the boundary will have to be changed to get to a ground state, the possible states being the mirror images of the earlier states reflected in the plane of the two spins in whose plane $\vec{S}_1$ has been reflected. However, only a few spins at the boundary need to be changed to go from one ground state to another.

 To summarize, it is possible for classical Heisenberg spins with two spin $\vec{S}_i . \vec{S}_j$-type interactions to have a ground state manifold which consists of disconnected pieces. Few-spin clusters as well as translationally invariant lattices were constructed to demonstrate this. The lattices discussed show a sharp transition, as the interaction strengths are varied with respect to each other, from a phase in which the ground state manifold is connected and continuous, to a phase in which it has an exponential number of disconnected pieces in it. An analogous transition was also observed for the six-spin cluster. The spin-half version of some of these lattice models is likely to show chiral spin liquid behavior.


\section*{Acknowledgements}

 I thank Professor Deepak Dhar for his guidance and encouragement through this work. I thank Professor John Chalker for suggesting the problem and for his comments. I also want to thank Professor G. Baskaran for his comments and feedback. I thank Council of Scientific and Industrial Research of the Government of India for financial support through Shyama Prasad Mukherjee Fellowship.




\begin{thebibliography}{05}


\bibitem{label1}
P.W.Anderson, Basic notions of Condensed Matter Physics, Westview Press, Colorado, 1984


\bibitem{label2}
M.Mezard, G.Parisi, M.A.Virasoro, Spin Glass Theory and Beyond, World Scientific, 1987

\bibitem{label3}
A.P.Ramirez, Annu.Rev.Mater.Sci., 24, 453 (1994)


\bibitem{label4}
L. Gammaitoni, P. Hänggi, P. Jung and F. Marchesoni, Rev. Mod. Phys. 70, 223-287 (1998)

\bibitem{label5}
See, for instance, theorem 1 in S.Chandra, Phys. Rev. E, 77, 021125 (2008)

\end{thebibliography}
\end{document}